\documentclass[10pt]{article}
\usepackage{amssymb,amsfonts,amsmath,latexsym}
\newcommand{\be}{\begin{equation}}
\newcommand{\ee}{\end{equation}}
\newcommand{\bea}{\begin{eqnarray}}
\newcommand{\eea}{\end{eqnarray}}
\newcommand{\ba}{\begin{array}}
\newcommand{\ea}{\end{array}}
\newcommand{\bt}{\begin{tabular}}
\newcommand{\et}{\end{tabular}}

\newcommand{\fr}{\frac}
\newcommand{\ci}{\cite}
\newcommand{\cl}{\centerline}
\newcommand{\bs}{\bigskip}

\newcommand{\en}{\eqno}

\newcommand{\bbib}{\begin{thebibliography}}
\newcommand{\ebib}{\end{thebibliography}}

\begin{document}

\cl{\bf ON EFFECTIVE CONDUCTIVITY }
\cl{\bf OF FLAT RANDOM TWO-PHASE MODELS}
\bs

\cl{\bf S.A.Bulgadaev \footnote{e-mail: sabul@dio.ru}}

\bs
\cl{Landau Institute for Theoretical Physics}
\cl{Chernogolovka, Moscow Region, Russia, 142432}

\bs

\begin{quote}
\footnotesize{
An approximate functional equation for
effective conductivity $\sigma_{eff}$ of systems with a finite maximal
scale of inhomogeneities is deduced. An exact solution of this equation is
found and its physical meaning is discussed.
A two-phase randomly inhomogeneous model is constructed by a hierarchical
method and its effective conductivity at arbitrary phase concentrations
is found in the mean-field-like approximation. These expressions satisfy
all necessary symmetries, reproduce the known formulas for $\sigma_{eff}$
in weakly inhomogeneous case and coincide with two recently found partial
solutions of the duality relation. It means that $\sigma_{eff}$ even of the
two-phase randomly inhomogeneous system may be a nonuniversal function and
can depend on some details of the structure of the inhomogeneous regions.
The percolation limit is briefly discussed.}
\end{quote}

\vspace{0.4cm}
\cl{PACS: 73.61.-r, 75.70.Ak}

\bs

The electrical transport properties of the classical inhomogeneous systems have
an important practical interest. The simplest problem in this region is
a finding of the effective conductivity $\sigma_{eff}$ of an isotropic
inhomogeneous (randomly or regularly) two-phase system, which is a mixture
of two phases with different conductivities  $\sigma_i \, (i = 1,2).$
Despite of its relative simplicity only a few general results have
been obtained so far.
In case of weakly inhomogeneous isotropic medium \ci{1}
$$
\sigma_{eff} = \langle \sigma \rangle \left(1 -
\fr{\langle \sigma^2 \rangle - {\langle \sigma \rangle}^2}
{2{\langle \sigma \rangle}^2}\right).
\en(1)
$$
For two-phase system
$\langle \sigma \rangle = x \sigma_1 + (1-x) \sigma_2, \;
\langle \sigma^2 \rangle - \langle \sigma \rangle^2 = 4x(1-x)(\sigma_-)^2,$
where $x$ is a concentration of the first phase,
$\sigma_- = (\sigma_1 - \sigma_2)/2$,
and (1) takes the form
$$
\sigma_{eff} =
\sigma_+\left(1 + 2\epsilon z - (1-4\epsilon^2) z^2/2\right),
\en(1')
$$
where $\sigma_+ = (\sigma_1 + \sigma_2)/2, \;
z = \sigma_-/\sigma_+ \; (-1 \le z \le 1), \; \epsilon =x-1/2.$

Another general formula is the dilute limit of the Maxwell--Garnett
formula \ci{1}
$$
\sigma_{eff} = \sigma_1 \left(1 -
2(1-x)z \right),
\en(2)
$$
where $1-x \ll 1$ is a small concentration of the second phase and a round
form of the inclusions of this phase is suggested.

Then, the exact
Keller -- Dykhne formula  for $\sigma_{eff}$ in systems with equal
concentrations of both phases and the exact dual relation, connecting
effective conductivities at adjoint concentrations $x$ and $1-x,$ have been
obtained \ci{2,3}
$$
\sigma_{eff}(x,\sigma_1,\sigma_2) \sigma_{eff}(1-x,\sigma_1,\sigma_2) =
\sigma_1 \sigma_2
\en(3)
$$
The exact Keller -- Dykhne formula follows from (3) at $x = x_c = 1/2$
$$
\sigma_{eff}(x_c,\sigma_1,\sigma_2) =
\sqrt{\sigma_1 \sigma_2}.
\en(4)
$$
This formula is very simple and universal, because it does not depend on
inhomogeneous structure and takes place for regular inhomogeneous
two-phase systems \ci{4,5} as well as for slightly nonregular inhomogeneous
systems \ci{5}.

Of course, a formula for the effective conductivity at arbitrary phase
concentrations has the main interest in this problem.
One such approximate formula for $\sigma_{eff}$ has been obtained many years
ago in the so called effective medium (EM) approximation \ci{6}
$$
\sigma_{eff}(\epsilon,\{\sigma\}) = \sigma_+ \left( 2\epsilon z +
\sqrt{1-z^2 + 4(\epsilon z)^2}\right).
\en(5)
$$
where $\{\sigma\} = (\sigma_1, \sigma_2).$
Though it corresponds to the weakly inhomogeneous case it turns out
to be a good approximation, when $\sigma_i \ne 0$ \ci{7}.

It was shown recently that the duality relation (3) together with boundary
conditions and some assumptions about the properties of $\sigma_{eff}$ allow
to find an explicit form of $\sigma_{eff}$ at arbitrary $x$
and two such expressions were found \ci{8,9}.
In this paper we will represent two randomly inhomogeneous models, having
their effective conductivities just of these two forms.
Another important question appears naturally in this problem:
is a formula for the effective conductivity
of randomly inhomogeneous two-phase systems universal or it can depend on the
structure of the inhomogeneities of the system? The strong arguments in favor
of the nonuniversality of the effective conductivity in case of three-phase
regular inhomogeneous systems were derived in the paper \ci{10}.
Our results for $\sigma_{eff}$ demonstrate that a formula for the
effective conductivity may be nonuniversal even in the two-phase case.
At the end of the paper we will briefly discuss some peculiarities
of the percolation limit.

We start our investigation with a general discussion of the averaging
procedure for obtaining $\sigma_{eff}$.
It is easy to see that the effective conductivity will depend
on a scale $l$ of a region over which an averaging is done.
This takes place
due to the possible existence of different characteristic scales in the
inhomogeneous medium. In the most general case there will be a whole spectrum
of these characteristic scales. This spectrum can be very different: from
discrete finite till continuous infinite, and is defined by the
structure of the inhomogeneities of the system.
For this reason it can depend on the phase concentration $x.$
Suppose, for the simplicity, that the randomly inhomogeneous structure of
our system has the scale spectrum with a maximal scale $l_m(x),$ which is
finite for all $x$ in the region $1 \ge x > 1/2$ (or $1-x$ in the region
$0 \le 1-x < 1/2$). Let us assume that we know an exact formula for
$\sigma_{eff}(x,\{\sigma\})$ of this system, which is applicable from
scales $l > l_m.$
It means that this formula for $\sigma_{eff}(x,\{\sigma\})$ takes place after
the averaging over regions with a mean size $l \gtrsim l_{m}$ and does not
change for all larger scales $l \gg l_m.$
Now consider a square lattice
with the squares of length $l_L \gg l_{m}$ and suppose that they have the
effective conductivities corresponding to different values of the
concentrations $x_1$ and $x_2$ with equal probabilities $p = 1/2$ (see Fig.1)

\begin{figure}
\begin{picture}(250,120)
\put(45,20){\line(0,1){50}}
\put(45,20){\line(1,0){50}}
\put(95,20){\line(0,1){50}}
\put(45,70){\line(1,0){50}}
\put(55,30){\circle*{15}}
\put(60,55){\circle*{10}}
\put(75,60){\circle*{20}}
\put(85,40){\circle*{10}}
\put(65,45){\circle*{5}}
\put(65,5){(a)}
\put(150,0){%
\begin{picture}(100,50)%
\multiput(75,20)(10,0){11}%
{\line(0,1){50}}
\multiput(75,20)(0,10){6}%
{\line(1,0){100}}
\multiput(77.5,21)(0,20){3}{1}
\multiput(77.5,31)(0,20){2}{2}
\multiput(87.5,31)(0,20){2}{1}
\multiput(87.5,21)(0,20){3}{2}
\multiput(97.5,21)(0,20){3}{1}
\multiput(97.5,31)(0,20){2}{2}
\multiput(107.5,31)(0,10){3}{2}
\multiput(107.5,21)(0,40){2}{1}
\multiput(117.5,51)(0,10){2}{2}
\multiput(117.5,21)(0,10){3}{1}
\multiput(127.5,21)(0,10){5}{2}
\multiput(137.5,21)(0,20){3}{1}
\multiput(137.5,31)(0,20){2}{2}
\multiput(147.5,21)(0,10){5}{1}
\multiput(157.5,31)(0,10){2}{2}
\multiput(157.5,21)(0,30){2}{1}
\multiput(167.5,31)(0,20){2}{2}
\multiput(167.5,21)(0,20){2}{1}
\put(167.5,61){2}
\put(157.5,61){2}
\put(120,5){(b)}
\end{picture}}
\end{picture}

{\small Fig.1. a) An elementary square of the model
with round inclusions, b) a lattice of the model; the numbers 1,2 denotes
squares with the corresponding concentrations.}
\end{figure}

After the averaging over the scales $l \gtrsim l_L$ one must obtain on much
larger scales $l \gg l_{L}$ the same effective conductivity,
but corresponding to another concentration $x=(x_1 + x_2)/2.$
This is possible due to the similar random structure of different squares
and due to the conjectures that (1) in this model there are only two important
scales: a maximum of the maximal characteristic scales (we suppose here
that $l_1 \sim l_2 \sim l_m(x) $, where $l_m(x_i) \equiv l_i \quad (i=1,2)$)
and the lattice square size $l_L \gg l_i$, (2) the averaging procedures over
these scales do not correlate (or weakly correlate) between themselves.
Thus, for a compatibility,
all concentrations must be out of small region around critical concentration
$x_c,$ where $l_i$ or $l_m(x)$ can be very large. We will call further
this set of the conjectures the finite maximal scale averaging approximation
(FMSA approximation). It can be considered as some nontrivial modification
of the EM approximation and can be implemented for systems with compact
inhomogeneous inclusions with finite $l_m$.

From the other side the effective conductivity
on scales $l \gg l_L$ must be determined  by the universal
Keller -- Dykhne formula (4). Thus we obtain the next functional equation
for the effective conductivity, connecting $\sigma_{eff}(x,\{\sigma\})$ at
different concentrations,
$$
\sigma_{eff}(x,\{\sigma\}) = \sqrt{\sigma_{eff}(x_1,\{\sigma\})
\sigma_{eff}(x_2,\{\sigma\})},\quad
x=(x_1+x_2)/2.
\en(6)
$$
It must be supplemented by the boundary conditions
$$
\sigma_{eff}(1, \{\sigma\}) = \sigma_1, \quad
\sigma_{eff}(0, \{\sigma\}) = \sigma_2,
\en(6')
$$
The equation (6) can be considered as a generalization of the duality
relation (3), the latter being a particular case of (6) at $x_1 + x_2 = 1.$
It follows from (6) that for $z \ne 1$ and due to the exactness of the
duality relation  it really works at all concentrations $x$
except maybe of small region near $x \ge x_c$ and $\sigma_2 = 0 \; (z=1)$
(see below a discussion of the percolation limit).
It is easy to see also that the approximate
formula (2) for $\sigma_{eff}(x,\{\sigma\})$
satisfies equation (6). Moreover, one can find an exact solution of this
equation. It has an exponential form
with a linear function of $x$ (or $\epsilon$)
$$
\sigma_{eff}(x,\{\sigma\}) = \sigma_{1} \exp(ax + b),
\en(7)
$$
where the constants $a,b$ can be determined from the boundary conditions
$$
a = - b, \quad \exp b = \sigma_2/\sigma_1.
\en(7')
$$
Substituting these coefficients into (7) one obtains
$$
\sigma_{eff}(x,\{\sigma\}) = \sigma_{1} \left(\sigma_2/\sigma_1\right)^{(1-x)}
= {\sigma_1}^x {\sigma_2}^{1-x}.
\en(8)
$$
The solution (8) satisfies all required symmetry relations and
exactly coincides with the case (a) from \ci{8,9}.

It is interesting to note that the form (8) means that
in the FMSA approximation one has effectively the averaging  of $\ln \sigma$
since it can be represented as
$$
\ln \sigma_{eff} = \langle \ln \sigma \rangle =
x \ln \sigma_1 + (1-x)\ln \sigma_2.
\en(9)
$$
This was noted already in \ci{3} for the case of equal concentrations
$x=1/2.$ The analogous result has been obtained later for random system in
the theory of two-dimensional weak localization \ci{11}.
One can check that (8) reproduces
in the weakly inhomogeneous limit the universal Landau -- Lifshitz expression
(1). In the low concentration limit of the second phase it gives
$$
\sigma_{eff}(x,z) = \sigma_1(1 + (1-x)\log \fr{1-z}{1+z} + ...),
\quad 1-x \ll 1,
\en(10)
$$
what coincides with (2) in the weakly inhomogeneous case.
Note that the expansion (10) contains the coefficients diverging in the limit
$|z| \to 1.$  Such behaviour of the coefficients denote the existence of a
possible singularity in this limit, where the FMSA approximation is not
applicable (see below a discussion of this percolation limit).

Now we will construct  a hierarchical model of flat isotropic randomly
inhomogeneous two-phase system, using the composite method introduced above,
and find its effective conductivity $\sigma_{eff}(x,\{\sigma\}).$

Let us consider a simple square lattice with the squares consisting of a
random layered mixture of two
conducting phase with constant conductivities $\sigma_i \; (i = 1,2)$ and
the corresponding concentrations $x$ and $1-x.$  A schematic picture of such
square is given in Fig.2.

\begin{figure}
\begin{picture}(250,120)
\put(50,20){\line(1,0){50}}
\put(50,20){\line(0,1){50}}
\put(100,20){\line(0,1){50}}
\put(50,70){\line(1,0){50}}
\put(60,20){\line(0,1){50}}
\put(63,20){\line(0,1){50}}
\put(80,20){\line(0,1){50}}
\put(90,20){\line(0,1){50}}
\multiput(66.5,45)(5,0){3}{\circle*{1}}
\multiput(61.5,23)(0,3){16}{\circle*{2}}
\multiput(83.3,23)(0,3){16}{\circle*{2}}
\multiput(86.6,23)(0,3){16}{\circle*{2}}
\put(65,5){(a)}
\put(150,0){%
\begin{picture}(100,50)%
\multiput(75,20)(10,0){11}%
{\line(0,1){50}}
\multiput(75,20)(0,10){6}%
{\line(1,0){100}}
\put(120,5){(b)}
\multiput(78,25)(0,20){3}{\line(1,0){3}}
\multiput(80,33)(0,20){2}{\line(0,1){3}}
\multiput(88,35)(0,20){2}{\line(1,0){3}}
\multiput(90,23)(0,20){3}{\line(0,1){3}}
\multiput(98,25)(0,20){3}{\line(1,0){3}}
\multiput(100,33)(0,20){2}{\line(0,1){3}}
\multiput(110,33)(0,10){3}{\line(0,1){3}}
\multiput(108,25)(0,40){2}{\line(1,0){3}}
\multiput(120,53)(0,10){2}{\line(0,1){3}}
\multiput(118,25)(0,10){3}{\line(1,0){3}}
\multiput(130,23)(0,10){5}{\line(0,1){3}}
\multiput(138,25)(0,20){3}{\line(1,0){3}}
\multiput(140,33)(0,20){2}{\line(0,1){3}}
\multiput(148,25)(0,10){5}{\line(1,0){3}}
\multiput(160,33)(0,10){2}{\line(0,1){3}}
\multiput(158,25)(0,30){2}{\line(1,0){3}}
\multiput(170,33)(0,20){2}{\line(0,1){3}}
\multiput(168,25)(0,20){2}{\line(1,0){3}}
\put(170,63){\line(0,1){3}}
\put(160,63){\line(0,1){3}}
\end{picture}}
\end{picture}

{\small Fig.2. a) An elementary square of the model with a
vertical orientation, the dotted regions denote layers of the second phase;
b)  a lattice of the model, the small lines on the
squares denote their orientations.}
\end{figure}

The layered structure
of the squares means that the squares have some preferred direction,
for example
along the layers. Let us suppose that the directions of different squares
are randomly oriented (parallely or perpendicularly) relatively to the
external electric field, which is directed along $x$ axis.
In order for system to be isotropic the probabilities of the parallel and
perpendicular orientations of squares must be equal or (what is the same)
the concentrations of the squares with different orientations must be equal
$p_{||} = p_{\perp} = 1/2.$

This structure can appear,
for example, on the small macroscopic scales, when a random medium is formed
as a result of the stirring of the two-phase mixture.
The corresponding averaged parallel and
perpendicular (or serial) conductivities of squares $\sigma_{||}(x)$ and
$\sigma_{\perp}(x)$ are defined by the following formulas
$$
\sigma_{||}(x) = x \sigma_1 + (1-x) \sigma_2 = \sigma_+ (1+ 2\epsilon z),
$$
$$
\sigma_{\perp}(x) = \left(\fr{x}{\sigma_1}+\fr{1-x}{\sigma_2}\right)^{-1}
= \sigma_+ \fr{1-z^2}{1-2\epsilon z}.
\en(11)
$$
Thus we have obtained the hierarchical representation of random medium
(in this case a two-level one). On the first level it consists  from
some regions (two different squares) of the random mixture of the two
layered conducting phases with different conductivities
$\sigma_1$ and $\sigma_2$ and arbitrary concentration. On the second level
this medium is represented as a random parquet constructed from two such
squares with different conductivities $\sigma_{||}$ and $\sigma_{\perp}$,
depending nontrivially on concentration of the initial conducting phases,
and randomly distributed with the same probabilities $p_i = 1/2$ (Fig.2).
This representation allows us to divide the averaging process into two
steps (firstly averaging over each square and then averaging over the lattice
of squares) and implement on the second step the exact formula (4).
This can be considered as some modification of the FMSA approximation.
As a result one obtains for the effective conductivity of the introduced
random two-phase model the following formula, which is applicable for
arbitrary concentration
$$
\sigma_{eff}(\epsilon,\{\sigma\}) = \sigma_+ f(\epsilon, z),\quad
f(\epsilon, z) = \sqrt{1-z^2}
\left[\fr{1 + 2\epsilon z}{1 - 2\epsilon z}\right]^{1/2},
\en(12)
$$
This function has all necessary properties, satisfies equation (2) and
coincides with the second form from \ci{8,9}.

It is interesting to compare this formula with the known general formulas.

(a) In case of small concentration of the first phase $x \ll 1$ one gets
$$
\sigma_{eff}(x,\{\sigma\}) \simeq  \sigma_2 \left(1 + \fr{2xz}{1-z^2}\right).
\en(13)
$$
It follows from (13) that an addition of small part of the first  higher
conducting phase increases an effective conductivity of the system as it
should be.

(b) In the opposite case of small concentration of the second phase
$1 - x  \ll 1$ one obtains
$$
\sigma_{eff}(x,\{\sigma\}) \simeq  \sigma_1 \left(1 - \fr{2(1-x)z}{1-z^2}\right),
\en(14)
$$
i.e. an addition of the phase with smaller conductivity decreases
$\sigma_{eff}.$ It is worth to note that both these expressions
for arbitrary values of the conductivities $\sigma_1$ and $\sigma_2$
differ from equation (2) and coincide with it only in the weakly inhomogeneous
case $z \ll 1$.
It must be not surprising because  a form of the inclusions of the second
phase in this model has completely different, layered, structure.
In the low concentration expansion one can  see again that the divergencies
appear in the limit $|z| \to 1.$

(c) In case of almost equal phase concentrations
$x = 1/2 + \epsilon,$ $\epsilon \ll 1$ one obtains
$$
\sigma_{eff}(\epsilon,\{\sigma\}) \simeq
\sigma_+ \sqrt{1-z^2} \left(1 + 2\epsilon z\right).
\en(15)
$$
The Keller -- Dykhne formula (3) is reproduced for equal concentrations.

One must note that at the same time the formula (12) does not satisfy the
equation (6) except of the trivial case $x_1 = x_2.$

For a comparison of the different expressions of the effective conductivity
 the plots of $f(\epsilon,z)$ in
the EM approximation, in FMSA approximation and of the
hierarchical model in FMSA-like approximation were constructed (see \ci{8,9}).
It follows from these plots  that all derived
above formulas for $\sigma_{eff}$, despite of their various functional forms,
differ from each other weakly  for
$z \lesssim 0,8$ due to very restrictive boundary conditions (6') and
the exact Keller-Dykhne value. This range of $z$ corresponds approximately
to the ratio $\sigma_2/\sigma_1 \sim 10^{-1}.$ For the smaller ratios a
difference between these functions become distinguishable.

Now let us consider in the more details the derived formulas for
$\sigma_{eff}(\epsilon,\{\sigma\})$
in case when $\sigma_2 \to 0 (z \to 1).$ It is clear that for regularly
inhomogeneous medium one can always construct such distribution of the
conducting phase that  $\sigma_{eff}(\epsilon,1)$ will differ from zero for all
$1/2 \ge \epsilon > -1/2$.
But in the case of randomly inhomogeneous medium the limit $\sigma_2 \to 0$
is equivalent to the well known percolation problem [12,13]. In terms
of $z$ it corresponds to the limit $z \to 1$ and is also similar to the
superconducting limit $\sigma_1 \to \infty$.
Strictly speaking,  an implementation of the duality
transformation (3) is not obvious in this case.
However, if one supposes that the dual symmetry relation (3) fulfills in this
limit too due to a continuity then it follows from (3) that
$$
\sigma_{eff}(\epsilon) \sigma_{eff} (-\epsilon) = 0.
\en(16)
$$
The relation  (16) does not contradict to the known basic results
of the percolation theory that
$\sigma_{eff}(\epsilon) = 0$ for $\epsilon \le 0$ and
$\sigma_{eff}(\epsilon) \ne 0$ for $\epsilon > 0$.
Moreover one can show that in this case
$$
\sigma_{eff}(\epsilon) =
\left\{\begin{array}{rc}
0, & \epsilon \le 0,\\
2\sigma_a, &  \epsilon > 0,\\
\end{array}\right.
\en(17)
$$
where $\sigma_a$ is the odd part of $\sigma_{eff}(\epsilon).$
It means that a behaviour of  $\sigma_{eff}(\epsilon)$ in the
percolation theory is completely determined by its odd part.
It is known from the experimental and numerical results that in the
percolation limit the effective conductivity $\sigma_{eff}$ have a
nonanalytical behaviour near the percolation edge $x_c = 1/2$ (or at small
$\epsilon > 0$)
$$
\sigma_{eff}(\epsilon) \sim \sigma_1 (x-x_c)^{t} \sim \sigma_1 \epsilon^t,
\en(18)
$$
where a critical exponent of the conductivity $t$ is slightly above 1
and can be represented in the form $t=1+\delta.$
Since the  values of this exponent found by the numerical calculations
are confined to be in the interval (1,10 -- 1,4) [13], then
$\delta$ have to be small and belongs to the interval (0.1 -- 0.4).
It follows from (17) that the same behaviour must have $\sigma_a$.
It means that at small $\epsilon$ there is
some crossover  on $z$ under $z \to 1$ from a regular (analytical)
behaviour to a singular one. At the moment an exact form of this crossover
is unknown.

From the formulas obtained above,
one gets always  $\sigma_{eff} \to 0$ in the limit  $\sigma_2 \to 0$, except
the region near $x=1.$
It means that all these formulas, obtained in FMSA approximation,
are not valid in the limit
$\sigma_2 \to 0.$ This is confirmed by the appearance of the divergencies
in the limit $z \to 1.$
This fact is a consequence of the made approximation.
For example, in case of the model of the layered squares this is due to the
"closing" (or "locking") effect of the layered structure
in the adopted approximation in the limit $\sigma_2 \to 0.$ In order
to obtain a finite conductivity in this model above threshold concentration
$x_c$ one needs to take into account the correlations between adjacent squares.
It is easy to show that near the threshold an effective conductivity is
determined by random conducting clusters formed out of the crossing
random layers from neighbouring elementary squares. As is well known,
the mean size of these clusters
diverges near the percolation threshold \ci{12,13}
and for this reason the FMSA approximation cannot be
applicable for the description of $\sigma_{eff}$ in the region $z\to 1$ and
$x \le 1/2.$
It follows from our results that EM approximation
overestimates $\sigma_{eff}$, whereas both other formulas underestimate it
in this region. We hope to investigate the percolation limit in detail
in the subsequent papers.

Thus, though both formulae for the effective
conductivity obtained above have the various functional forms, they
satisfy all symmetries, including the dual symmetry and all necessary
inequalities, and reproduce the general formulae for $\sigma_{eff}$
in the weakly inhomogeneous case. These results allow us to make a
conjecture that $\sigma_{eff}$ even of the
two-phase randomly inhomogeneous systems may be a nonuniversal function and
can depend on some details of the structure of the randomly inhomogeneous
regions.
The obtained formulae can be considered as the regular ones, since
they are applicable only for systems with $\sigma_i \ne 0,$
when there are no singularities connected with a percolation problem.
The introduced composite method of the construction of the model
random medium
can be generalized on the heterophase systems with arbitrary number of phases
and on the other ways of determination of the effective
intermediate conducting boxes. It can be done for various types of boxes
as well as for different numbers of the possible types of the boxes.
It is clear that then one will have to use instead of (4) another formula.
The constructed models and obtained expressions can be used for the modelling
and description of some real composite systems.

The author thanks referees for useful remarks.
This work was supported by RFBR grants \# 2044.2003.2 and 02-02-16403.


\bbib{20}
\bibitem{1} Landau L.D., Lifshitz E.M.,{\it Electrodynamics of condensed media},
Moscow, 1982 (in Russian).
\bibitem{2} Keller J.B., {\it J.Math.Phys.}, {\bf 5}  (1964) 548.
\bibitem{3} Dykhne A.M., {\it ZhETF}, {\bf 59}  (1970) 110 (in Russian).
\bibitem{4} Emetz Yu.P., {\it  ZhETF}, {\bf 96}  (1989) 701 (in Russian).
\bibitem{5} Ovchinnikov Yu.N., Dyugaev A., {\it ZhETF}, {\bf 117} (2000) 1013;
Balagurov B.Ya., {\it ZhETF}, {\bf 117} (2000)  1561,
{\bf 118} (2001)  665 (in Russian).
\bibitem{6}  Bruggeman D.A.G., {\it Ann.Physik}, {\bf 24}(1935)  636;
 Landauer R., {\it J.Appl.Phys.}, {\bf 23} (1952)  779.

\bibitem{7} Kirkpatrick S., {\it Phys.Rev.Lett.}, {\bf 27} (1971) 1722.
\bibitem{8} Bulgadaev S.A., Preprint ITP 02-12-02 (2002), cond-mat/0212104.
\bibitem{9} Bulgadaev S.A., {\it Phys.Lett.}, {\bf A313} (2003) 106.
\bibitem{10}
 Fel L.G., Machavariani V.Sh., Khalatnikov I.M. and Bergman D.J.,
{\it J.Phys.}, {\bf A33} (2000) 6669.
\bibitem{11} Anderson P.W., Thouless D.J., Abrahams E. and Fisher D.S.,
{\it Phys.Rev.}, {\bf B22} (1980) 3519.
\bibitem{12}  Kirkpatrick S., {\it Rev.Mod.Phys.}, {\bf 45}  (1973) 574.
\bibitem{13} Shklovskii B.I., Efros A.L., {\it Electronic Properties of
Doped Semiconductors,} Vol.{\bf 45}, Springer Series in Solid State Sciences,
(Springer Verlag, Berlin) 1984.

\ebib
\end{document}